\documentclass[aps, twocolumn, nofootinbib, showpacs, prl]{revtex4}
\usepackage{graphicx}
\usepackage{amsfonts}
\usepackage{amssymb}
\usepackage{amsbsy}
\usepackage{amsmath}
\usepackage{latexsym}
\usepackage{natbib}
\usepackage{bm}
\usepackage{color}
\usepackage{float}

\begin{document}
\title{On the Polynomial Degeneracy of Ricci Invariants and Spacetime Singularity}

\author{Soumya Chakrabarti}
\email{soumya.chakrabarti@vit.ac.in}

\affiliation{School of Advanced Sciences, Vellore Institute of Technology, Vellore, Tiruvalam Rd, Katpadi, Tamil Nadu 632014, India}


\date{\today}

\begin{abstract}
We explore the connection of a general relativistic matter-energy momentum tensor with the polynomial degeneracies of higher order curvature invariants defined in Riemannian geometry. The degeneracies enforce additional constraints on the energy-momentum tensor components. Due to these constraints the formation of a curvature singularity, for instance during a gravitational collapse can no longer be treated as inevitable. We find that there can be a formation of singularity iff the interior fluid evolves into $(i)$ a pressure-less dust, $(ii)$ an isotropic sphere or $(iii)$ a distribution with negative pressure.      
\end{abstract}

\maketitle

The concept of a spacetime singularity emerges naturally in General Theory of Relativity (GR). The theory provides a distinct framework where spacetime curvature governs the movement of objects and energy-momentum distributions. This evolution is encapsulated through a set of coupled non-linear field equations. One of the most notable solutions of these equations is the Schwarzschild solution, which characterizes a unique metric tensor describing static, spherically symmetric vacuum \cite{sch, schsup}. It also features a singularity at its center, i.e., at $r = 0$. This can be interpreted as an infinitesimally small region where matter/energy is confined, resulting in an infinite curvature of spacetime. In technical words singularities are an incompleteness of geodesics and a general breakdown of physical predictability. The presence of these singular points may seem problematic in a fundamental theory of interaction. However, in GR a formation of physical singularities is permitted and commonly associated with black holes, characterized by gravitational forces so intense that not even photons can escape their grasp. \\

Physics near singularities can be a fascinating yet challenging subject to talk about. In order to preserve the conventional understanding of nature at an asymptotic limit, usually a notion of horizon is implemented. For most of the singular solutions of the field equations in GR, there is also an associated formation of trapped surface \cite{bh1} or an event horizon. This surface effectively isolates the singularity from any external observers and creates a Black Hole. However, counter-examples can also be found in the form of solutions with no formation of horizon \cite{ns1, ns2}, known as Naked Singularities. There are also reasonable examples of non-singular solutions such as regular Black Holes \cite{regular, regular1} or Wormholes \cite{sc} where there is no singularity at $r = 0$. Apart from a few formulations of censorship \textit{conjecture} \cite{censor}, there is nothing much to go on and make an educated prediction about the formation of singularities based on the initial configurations. In this article we explore the role of scalar curvature invariants and the flow of their gradient fields in this matter. This area of study has received some interests from relativists over the years \cite{flow} and provided different possibilities related to the geometric features of a spacetime metric itself. We aim to leverage these insights and predict (without solving the field equations) whether a given spacetime geometry permits the formation of singularities. This involves two key steps : (i) identifying algebraically independent curvature invariants and (iii) correlating their interdependence with the components of a generic energy-momentum tensor. The approach is independent of the theory of gravity under consideration and relies solely on the geometric properties of curvature tensors. Nonetheless, to establish a meaningful connection between these invariants and the energy-momentum components, we adopt the most well-established gravitational framework : GR. \\

Any geometric contraction can not produce an algebraically independent scalar invariant. Scalars that are functionally dependent but algebraically independent adhere to certain polynomial identities known as \textit{syzygies} \cite{syzy1}. The number of such invariants for a $d$-dimensional semi-Riemannian manifold can be expressed as
\begin{equation}\label{thomas}
\mathcal{N}(d,p)=\frac{d(d+1)((d+p)!)}{2d!p!}-\frac{(d+p+1)!}{(d-1)!(p+1)!}+d.
\end{equation}

A notable example is $\mathcal{N}(4,2) = 14$, representing the number of invariants without derivatives of the Riemann tensor (note that the construction of a scalar involves partial derivatives of the metric tensor up to order $p$). However, identifying a complete minimal set of invariants remains a challenging mathematical task, even for $d = 4$. This difficulty stems from the complexity of solving degeneracy equations, making the study of polynomial invariants an intriguing area of research \cite{syzy11, syzy111, syzy2}. While analyzing the dynamics permitted by a spacetime geometry, for instance, to look for conditions leading to a singularity formation, researchers often attempt to derive exact solutions to the field equations. These solutions frequently involve idealizations and lead to significant reductions in $\mathcal{N}$. As in the case of a Ricci-flat metric, the gradient fields of curvature invariants are highly constrained, reducing $\mathcal{N}(4,2)$ to just $1$ \cite{syzy3}. We argue that for a more realistic description, imposing simplifications or symmetries at the outset should be avoided, just to ensure that the mathematical properties remain independent of the observer four-velocity, preserving observer-independent gradient flows in the system. This in turn generates algebraically independent polynomial invariants and in this work, we focus on four such invariants including the Ricci scalar. The standard field equations of GR are 
\begin{equation}\label{einstein}
R^{\alpha}_{\beta}-\frac{1}{2}\delta^{\alpha}_{\beta}R+\Lambda \delta^{\alpha}_{\beta} = T^{\alpha}_{\beta}.
\end{equation}
$R^{\alpha}_{\beta}$ and $R$ denote the Ricci tensor and scalar respectively. $\Lambda$ can be kept as the cosmological constant while $T^{\alpha}_{\beta}$ and $T$ refer to the energy-momentum tensor and its trace. From this setup it is trivial to derive the first Ricci invariant or the Ricci scalar
\begin{equation}\label{ricci}
R = 4\Lambda - T.
\end{equation}

We note here, that the most natural process through which a general relativistic system can evolve towards a curvature singularity is a gravitational collapse. The context is that a massive astronomical object will eventually be depleted of its internal energy reserves and will be collapsing under an omnipresent gravitational force towards the centre, until a zero proper volume is reached. At this very moment the curvature invariants are expected to diverge. Traditionally, describing this process mathematically involves solving the non-linear field equations of the theory and checking for the divergence of the Ricci scalar. However, we want to ask a rather less-explored question : what constraints are already imposed on the potential formation of a singularity by the algebraically independent invariants beyond just the Ricci scalar and their polynomial degeneracies (especially when the well-known Ricci-flat assumption is not applied to the metric tensor)? In this connection, we first highlight three independent higher order invariants which are derived from combinations of the trace-free Ricci tensor and Weyl tensor \cite{syzy4}.

\begin{equation}\label{tracefree}
S^{\alpha}_{\beta} = R^{\alpha}_{\beta} -\frac{R}{4}\delta^{\alpha}_{\beta}.
\end{equation}
We can write by taking three contractions of the above equation
\begin{equation}\label{r1}
r_1 = S^{\alpha}_{\beta}S^{\beta}_{\alpha} ~,~ r_2 =S^{\alpha}_{\beta}S^{\gamma}_{\alpha}S_{\gamma}^{\beta} ~,~ r_3 =S^{\alpha}_{\beta}S^{\gamma}_{\alpha}S^{\delta}_{\gamma}S^{\beta}_{\delta}.
\end{equation}

We correlate these invariants with the energy-momentum tensor by using the Einsten field equation 
\begin{equation}\label{r1t}
r_1 = \frac{1}{4}(T^{\alpha}_{\beta}T^{\beta}_{\alpha}-\frac{T^2}{4}).
\end{equation}
Similar higher order expressions can easily be derived for $r_2$ and $r_3$. Using the Weyl Tensor, it is also possibe to write a fourth invariant as

\begin{equation}\label{w2}
w_{2} \sim C_{\alpha\beta\lambda\delta}C^{\alpha\beta\gamma\psi}C^{\lambda\delta}_{\gamma\psi}.
\end{equation} 

It can be proved \cite{syzy3, syzy4, syzy5} that not all of the higher order invariants are algebraically independent. Depending on the assumed background symmetry of the metric, in this case a spherical symmetry, there exist syzygies, which are fundamental identities connecting the associated invariants, such as $r_{1}, r_{2},.., w_{2}, w_{3},...$. For any theory of gravity linking the Einstein tensor to the stress-energy tensor, these syzygies should impose fundamental constraints on the components of the stress-energy tensor. In the case of simple energy-momentum distributions, such as a dust or a perfect fluid, the syzygies yield identities that do not necessitate additional restrictions. However, for more generalized distributions, such as a fluid with pressure anisotropy or heat flux, these constraints become significant. While the presence of algebraic restrictions have been briefly discussed in some studies, they have received little attention in the context of a singularity. For example, Narlikar and Karmarkar \cite{narlikar} demonstrated in their study of inequivalent spacetimes that for a spherically symmetric spacetime, only four invariants - namely the set  $\{R, r_1, r_2, w_1\}$ are truly independent. Further discussions on this topic can be found in the works of Koutras and McIntosh \cite{koutras}, Tipler, Clarke and Ellis \cite{tipler}. The challenge of identifying a complete set of invariants has also sparked considerable interest in the literature, for instance one can see the works of Carminati and McLenaghan \cite{syzy4}, Sneddon \cite{syzy11}, Zakhary and McIntosh \cite{syzy111}. Santosuosso et. al. \cite{syzy3} conducted an extensive study of invariants and their associated syzygies in spacetimes expressed as a product of a Lorentzian and a Riemannian manifold, described as
\begin{equation}
ds^2 = ds_{\Sigma_1}^2 (u,v) + C(x^\gamma)^2 ds_{\Sigma_2}^2(\theta,\phi),
\end{equation}
where $C(x^\gamma)^2$, the function coupling the $2$-spaces is constrained by a separability condition
\begin{equation}
C(x^\gamma)^2 = r(u,v)^2 w(\theta,\phi)^2.
\end{equation}
These are generally known as warped product spacetimes of class $B$ (Carot and Costa \cite{carot}, Nakahara \cite{nakahara}, Haddow and Carot \cite{haddow}). The choice of $sig(\Sigma_1)$ and $sig(\Sigma_2)$ can reproduce various classes of spacetime geometries which are commonly studied in GR. For example, $sig(\Sigma_1) = 0$ and $sig(\Sigma_2) = 2\epsilon$ ($\epsilon = \pm 1$) correspond to a class $B_1$ spacetime, which encompasses spherical, planar, and hyperbolic geometries. For a standard spherically symmetric spacetime $\{R, r_1, r_2, w_1\}$, the set of independent associated invariants satisfy an algebraic syzygy \cite{syzy3} written as
\begin{equation}
(-12r_{3}+7r_{1}^{2})^{3}-(12r_{2}^{2}-36r_{1}r_{3}+17r_{1}^{3})^{2}=0. \label{r33}
\end{equation}

To interpret the scenario physically, we note once again that Ricci invariants $r_n$ for $n > 3$ are not all independent. Additional syzygies for all $r_n$ with $n > 3$ can be constructed algorithmically. Explicit forms of these syzygies have been provided up to $r_10$ in literature \cite{syzy3}. More importantly, Ricci invariants are observer-independent. If we consider an observer with a tangent $4$-vector $u^{\alpha}$, the Ricci invariants do not include any $u^{\alpha}$ term. The energy-momentum tensor $T^{\alpha}_{\beta}$ can sometimes select a preferred $u^{\alpha}$ as in the case of perfect fluids, where it corresponds to the fluid streamlines. However, more general timelike $4$-velocities do not appear in the Ricci invariants. In a perfect fluid, Ricci invariants can be simplified into polynomials (up to quartic terms) involving the energy density and isotropic pressure. For an imperfect fluid the polynomials should carry additional informations. For example, during a gravitational collapse, gravity keeps on reinforcing itself and growing stronger as it feeds on its own intensity. This tendancy can cause the curvature scalar to exhibit sudden or asymptotic jumps towards divergence. For a spherically symmetric metric tensor, the algebraic identities between associated curvature scalar invariants must have a say on this question : under what condition such a jump towards divergence is allowed? Since the identities can be reformulated in terms of the components of energy-momentum tensor, we can reframe the question and ask : how should the matter distribution within a general relativistic sphere behave in order to achieve/avoid a singularity? \\

For a perfect fluid, combining the syzygy in Eq. (\ref{r33}) with Eqs. (\ref{r1}) and (\ref{w2}) a simple restriction on the energy-momentum tensor can be derived. A perfect fluid is usually realized by the energy-momentum tensor

\begin{equation}
T_{\alpha \beta} = (\rho + p) u_{\alpha} u_{\beta} + p g_{\alpha \beta},
\end{equation}

for which the Ricci invariants can be derived as
\begin{equation}\label{perfect}
r_{n} \sim (\rho + p)^{n+1}.
\end{equation}

Using Eq. (\ref{perfect}), it is straightforward to check that the syzygy in Eq. (\ref{r33}) is automatically satisfied. On the other hand, for an energy-momentum tensor incorporating anisotropic pressure and radial heat flux, written as
\begin{equation}\label{EM_gen}
T_{\alpha \beta} = (\rho_{e} + p_t) u_{\alpha} u_{\beta} + (p_{r} - p_t) n_{\alpha} n_{\beta} + p_t g_{\alpha\beta} + 2 u_{(\alpha}q_{\beta)},
\end{equation}
Eq. (\ref{r33}) produces an algebraic constraint
\begin{equation}
(q_{\alpha}q^{\alpha}-(n_{\alpha}q^{\alpha})^{2})^{2}(p_{re}-p_t)^{2} {\bf \it M} = 0. \label{P}
\end{equation}
The quantity ${\bf \it M}$ (written below) is a polynomial in $\rho_{e}, p_{r}, p_{t}, n_{\alpha} q^{\alpha}, q_{\alpha}q^{\alpha}$. If the system is isotropic this identity is automatically satisfied. If not, i.e., $p_{r} \neq p_{t} ~;~ q_{\alpha}q^{\alpha} \neq (n_{\alpha}q^{\alpha})^{2}$, Eq. (\ref{P}) reduces to 

\begin{widetext}
\begin{eqnarray}\label{mastereq}\nonumber
&&{\bf \it M} \equiv -p_{t}^{2}\rho_{e}^{4} - p_{r}^{2}\rho_{e}^{4} + 2p_{r}p_{t}\rho_{e}^{4} - 2p_{t}^{3}\rho_{e}^{3} - 4p_{t}(n_{\alpha}q^{\alpha})^2\rho_{e}^{3} + 4(n_{\alpha}q^{\alpha})^{2}p_{r}\rho_{e}^{3} + 2q_{\alpha}q^{\alpha}p_{t}\rho_{e}^{3} - 2q_{\alpha}q^{\alpha}p_{r}\rho_{e}^{3} - 2p_{r}^{3}\rho_{e}^{3} \\&&\nonumber
+ 2p_{r}^{2}p_{t}\rho_{e}^{3} + 2p_{r}p_{t}^{2}\rho_{e}^{3} -(q_{\alpha}q^{\alpha})^{2}\rho_{e}^{2} - 2p_{r}^{3}p_{t}\rho_{e}^{2} + 6(n_{\alpha}q^{\alpha})^{2}p_{r}^{2}\rho_{e}^{2} - p_{r}^{4}\rho_{e}^{2} + 8q_{\alpha}q^{\alpha}p_{t}^{2}\rho_{e}^{2} + 2q_{\alpha}q^{\alpha}p_{r}^{2}\rho_{e}^{2} - 2p_{r}p_{t}^{3}\rho_{e}^{2} \\&&\nonumber
- 6p_{t}^{2}(n_{\alpha}q^{\alpha})^{2}\rho_{e}^{2} - p_{t}^{4}\rho_{e}^{2} - 10q_{\alpha}q^{\alpha}p_{r}p_{t}\rho_{e}^{2} + 6p_{r}^{2}p_{t}^{2}\rho_{e}^{2} + 8(q_{\alpha}q^{\alpha})^{2}p_{r}\rho_{e} - 30p_{t}^{2}(n_{\alpha}q^{\alpha})^{2}p_{r}\rho_{e} + 2q_{\alpha}q^{\alpha}p_{t}^{3}\rho_{e} \\&&\nonumber
+ 6p_{t}^{3}(n_{\alpha}q^{\alpha})^{2}\rho_{e} + 8q_{\alpha}q^{\alpha}p_{r}^{3}\rho_{e} + 2p_{r}^{3}p_{t}^{2}\rho_{e} + 30p_{r}^{2}(n_{\alpha}q^{\alpha})^{2}p_{t}\rho_{e} + 2p_{t}^{3}p_{r}^{2}\rho_{e} + 18q_{\alpha}q^{\alpha}p_{t}(n_{\alpha}q^{\alpha})^{2}\rho_{e} - 6p_{r}^{3}(n_{\alpha}q^{\alpha})^{2}\rho_{e} \\&&\nonumber
- 2p_{r}p_{t}^{4}\rho_{e} + 10q_{\alpha}q^{\alpha}p_{t}^{2}p_{r}\rho_{e} - 10(q_{\alpha}q^{\alpha})^{2}p_{t}\rho_{e} - 2p_{t}p_{r}^{4}\rho_{e} - 18q_{\alpha}q^{\alpha}(n_{\alpha}q^{\alpha})^{2}p_{r}\rho_{e} - 20p_{r}^{2}q_{\alpha}q^{\alpha}p_{t}\rho_{e} + 2p_{r}^{2}p_{t}^{2}q_{\alpha}q^{\alpha} \\&&\nonumber
- 18q_{\alpha}q^{\alpha}p_{t}^{2}(n_{\alpha}q^{\alpha})^{2} + 10p_{r}^{3}(n_{\alpha}q^{\alpha})^{2}p_{t} - 10p_{t}^{3}(n_{\alpha}q^{\alpha})^{2}p_{r} - p_{t}^{4}p_{r}^{2} - (q_{\alpha}q^{\alpha})^{2}p_{t}^{2} - 4(n_{\alpha}q^{\alpha})^{2}p_{r}^{4} - 8p_{r}^{3}q_{\alpha}q^{\alpha}p_{t} \\&&\nonumber
+ 4q_{\alpha}q^{\alpha}p_{r}^{4} - 8p_{r}(q_{\alpha}q^{\alpha})^{2}p_{t} + 4(q_{\alpha}q^{\alpha})^{3} + 2p_{t}^{3}p_{r}q_{\alpha}q^{\alpha} + 2p_{t}^{3}p_{r}^{3} + 54(n_{\alpha}q^{\alpha})^{2}p_{r}q_{\alpha}q^{\alpha}p_{t} + 4p_{t}^{4}(n_{\alpha}q^{\alpha})^{2} \\&&
- 54(n_{\alpha}q^{\alpha})^{4}p_{r}p_{t} + 27p_{t}^{2}(n_{\alpha}q^{\alpha})^{4} + 27(n_{\alpha}q^{\alpha})^{4}p_{r}^{2} - 36q_{\alpha}q^{\alpha}(n_{\alpha}q^{\alpha})^{2}p_{r}^{2} - p_{r}^{4}p_{t}^{2} + 8(q_{\alpha}q^{\alpha})^{2}p_{r}^{2} = 0
\end{eqnarray}.
\end{widetext}

The first thing to note from Eq. (\ref{mastereq}) is that the evolution of energy-momentum tensor components such as density, pressure and heat flux can not be arbitrary. They can only evolve in a restricted manner satisfying this polynomial equation for all coordinate values of $r$ and $t$. Solving Eq. (\ref{mastereq}) in its original form is not straightforward and we proceed with a few reasonable approximations to pose our arguments. Near a spacetime singularity, properties of a spherical distribution of imperfect fluid should see through some radical changes within itelf. For example, near the end of a gravitational collapse, when the interior fluid distribution has radiated out almost entirely, one expect the density $\rho_{e}$ to dominate over all other energy-momentum tensor components. It is also possible to visualize the system after neglecting tangential pressure ($p_t = 0$). One might also define equation of state parameters for both radial and tangential pressure and analyze the equation. We will classify such approximations below and discuss the consequences. We also note that for the collapsing fluid to remain physical, it must obey four energy conditions. The most important is the Null Energy Condition which holds if for any null vector $T_{\mu\nu} k^{\mu} k^{\nu} \geq 0$. A weak energy condition dictates that for a non-spacelike vector $w^{\alpha}$, $T_{\alpha\beta} w^{\alpha} w^{\beta} \geq 0$. These two conditions makes sure that the energy density of the interior fluid remains non-negative. A dominant energy condition is satisfied if $- T_{\alpha\beta} w^{\beta}$ is timelike/null for a timelike vector $w^{\alpha}$. Effectively this conditions implies that the energy flow of matter has a speed less than the speed of light, for any observer. Finally, the strong energy condition can be enforced through $2 T_{\alpha\beta} w^{\alpha} w^{\beta} + T \geq 0$, where $w^{\alpha}$ is a timelike unit vector and $T$ is the trace of the energy momentum tensor. For an imperfect fluid these energy conditions can be written as algebraic inequalities \cite{energy}. For all the approximations of Eq. (\ref{mastereq}) discussed below, it is assumed that the energy conditions are satisfied. \\

{\bf Case I : Density domination}  \\
It is a possibility that during the end-state of a gravitational collapse most of the internal energy of the interior is either spent or already emitted as a radial heatflux. In such a case, we can keep only the linear order terms in $p_r$, $p_t$ and $q^{\alpha}$, ignoring any higher order terms. We also argue that if density of the collapsing sphere is much higher compared to these components, we can keep the highest two powers of $\rho_e$, namely $\rho_{e}^4$ and $\rho_{e}^3$. Under this approximation Eq. (\ref{mastereq}) can be written as
\begin{equation}
(p_r - p_t)\rho_{e}^4 + 2\rho_{e}^{3}(p_{r}^2 - p_{t}^{2}) + \rho_{e}^{3} \left\lbrace 2q_{\alpha}q^{\alpha} - 4(n_{\alpha}q^{\alpha})^{2} \right\rbrace = 0.
\end{equation}   

This can be re-written as
\begin{equation}\label{condition1}
\rho_{e} = \frac{4(n_{\alpha}q^{\alpha})^{2} - 2(p_{r}^2 - p_{t}^{2}) - 2q_{\alpha}q^{\alpha}}{(p_r - p_t)}.
\end{equation}

Eq. (\ref{condition1}) suggests that $\rho_{e}$ can not diverge unless there is a pressure isotropy. \\

{\bf Case II : Zero tangential pressure}  \\
We also consider a case where there is no tangential pressure of the collapsing fluid. We also represent $p_{r} = w \rho_e$ and note that due to the absence of $p_t$, there is always an in-built pressure anisotropy, unless $w = 0$. Under these conditions Eq. (\ref{mastereq}) becomes
\begin{eqnarray}\nonumber\label{condition2}
&& -\rho_{e}^{6}\Big \lbrace w^2 (1+w^2) \Big \rbrace + \rho_{e}^{4}\Big \lbrace (n_{\alpha}q^{\alpha})^2 2w(1-w)(2+2w \\&&\nonumber
+2w^2 +3w) - 2(q_{\alpha}q^{\alpha})w(1-w-4w^2 - 4w^3)\Big \rbrace \\&&\nonumber
-\rho_{e}^{2}\Big \lbrace (q_{\alpha}q^{\alpha})^2 (1-8w-8w^2) + 324(q_{\alpha}q^{\alpha})(n_{\alpha}q^{\alpha})^{2}\\&&
w(1+2w) - 27(n_{\alpha}q^{\alpha})^{4}w^{2}\Big \rbrace + 4(q_{\alpha}q^{\alpha})^{3} = 0.
\end{eqnarray}  

The above equation is cubic in $x = \rho_{e}^2$ and can be re-written in the form $ax^{3} + bx^{2} + c x + d = 0$. If $p$, $q$ and $r$ are the roots of such a cubic equation, they must obey the following,
\begin{eqnarray}\label{con1}
&& p + q + r = -\frac{b}{a}, \\&& \label{con2}
pq + qr + rp = \frac{c}{a}, \\&& \label{con3}
pqr = -\frac{d}{a}.
\end{eqnarray}

Comparing with Eq. (\ref{condition2}), we can write, for three roots $\rho_{1}^2$, $\rho_{2}^2$ and $\rho_{3}^2$,
\begin{widetext}
\begin{eqnarray}\label{condition3}
&&\left(\rho_{1}^2 + \rho_{2}^2 + \rho_{3}^2 \right) = \frac{\Big \lbrace (n_{\alpha}q^{\alpha})^2 2w(1-w)(2+2w+2w^2 +3w) - 2(q_{\alpha}q^{\alpha})w(1-w-4w^2 - 4w^3)\Big \rbrace}{\Big \lbrace w^2 (1+w^2) \Big\rbrace}, \\&&\label{condition4}
\left(\rho_{1}^2 \rho_{2}^2 + \rho_{2}^2 \rho_{3}^2 + \rho_{3}^2 \rho_{1}^2 \right) = \frac{\Big \lbrace (q_{\alpha}q^{\alpha})^2 (1-8w-8w^2) + 324(q_{\alpha}q^{\alpha})(n_{\alpha}q^{\alpha})^{2} w(1+2w) - 27(n_{\alpha}q^{\alpha})^{4}w^{2}\Big \rbrace}{\Big \lbrace w^2 (1+w^2) \Big\rbrace}, \\&&\label{condition5}
\left(\rho_{1}^2 \rho_{2}^2 \rho_{3}^2 \right) = \frac{4(q_{\alpha}q^{\alpha})^{3}}{w^2 (1+w^2)}
\end{eqnarray}
\end{widetext}

We find that Eqs. (\ref{condition3}), (\ref{condition4}) and (\ref{condition5}) suggest that $\rho_{e}$ can not diverge unless $w = 0$ or $w = -1$. $w = 0$ indicates that the radial pressure is also zero. This describes a dust-like behavior of the collapsing interior fluid. $w = -1$ indicates a negative pressure as in an effective dark energy-like behavior.  \\

{\bf Case III : End-state with non-zero pressure terms}  \\
In a scenario where both radial and tangential pressure are non-zero Eq. (\ref{condition2}) must be modified. In order to describe it in a mathematically approachable manner, we write the pressure components as $p_r = w_1 \rho_e$ and $p_t = w_2 \rho_e$. In this construct Eq. (\ref{mastereq}) can be simplified and re-ordered into

\begin{widetext}
\begin{eqnarray}\nonumber
&& \rho_{e}^{6} \Big[w_{2}^2 - w_{1}^2 + 2w_{1}w_{2} - 2w_{2}^3 + 2w_{1}^{2}w_{2} + 2w_{2}^{2}w_{1} - 2w_{1}^{3}w_{2} - w_{1}^{4} - 2w_{1}w_{2}^{3} - w_{2}^{4} + 6w_{1}^{2}w_{2}^{2} + 2w_{1}^{3}w_{2}^{2} + 2w_{2}^{3}w_{1}^{2} - 2w_{1}w_{2}^{4} \\&&\nonumber
- 2w_{2}w_{1}^{4} - w_{2}^{4}w_{1}^{2} + 2w_{1}^{3}w_{2}^{3} - w_{1}^{4}w_{2}^{2} - 2w_{1}^{3} \Big] + \rho_{e}^{4} \Big[-4\Big(n_{\alpha}q^{\alpha}\Big)^{2}w_{2} + 4\Big(n_{\alpha}q^{\alpha}\Big)^{2}w_{1} + 2\Big(q_{\alpha}q^{\alpha}\Big)w_{2} - 2\Big(q_{\alpha}q^{\alpha}\Big)w_{1} \\&&\nonumber
+ 6\Big(n_{\alpha}q^{\alpha}\Big)^{2}w_{1}^{2} + 8\Big(q_{\alpha}q^{\alpha}\Big)w_{2}^{2} + 4\Big(q_{\alpha}q^{\alpha}\Big)w_{1}^{2} - 6\Big(n_{\alpha}q^{\alpha}\Big)^{2}w_{2}^{2} - 10\Big(q_{\alpha}q^{\alpha}\Big)w_{1}w_{2} - 30\Big(n_{\alpha}q^{\alpha}\Big)^{2}w_{1}w_{2}^{2} + 2\Big(q_{\alpha}q^{\alpha}\Big)w_{2}^{3} \\&&\nonumber
+ 6\Big(n_{\alpha}q^{\alpha}\Big)^{2}w_{2}^{3} + 8\Big(q_{\alpha}q^{\alpha}\Big)w_{1}^{3} + 30\Big(n_{\alpha}q^{\alpha}\Big)^{2}w_{1}^{2}w_{2} - 6\Big(n_{\alpha}q^{\alpha}\Big)^{2}w_{1}^{3} + 10\Big(q_{\alpha}q^{\alpha}\Big)w_{2}^{2}w_{1} - 20\Big(q_{\alpha}q^{\alpha}\Big)w_{1}^{2}w_{2} + 10\Big(n_{\alpha}q^{\alpha}\Big)^{2}\\&&\nonumber
w_{2}w_{1}^{3} - 10\Big(n_{\alpha}q^{\alpha}\Big)^{2}w_{2}^{3}w_{1} - 4\Big(n_{\alpha}q^{\alpha}\Big)^{2}w_{1}^{4} - 8\Big(q_{\alpha}q^{\alpha}\Big)w_{1}^{3}w_{2} + 4\Big(q_{\alpha}q^{\alpha}\Big)w_{1}^{4} + 2\Big(q_{\alpha}q^{\alpha}\Big)w_{2}^{3}w_{1} + 4\Big(n_{\alpha}q^{\alpha}\Big)^{2}w_{2}^{4} \Big] \\&&\nonumber
+ \rho_{e}^{2} \Big[-\Big(q_{\alpha}q^{\alpha}\Big)^{2} + 8\Big(q_{\alpha}q^{\alpha}\Big)^{2}w_{1} + 18 \Big(q_{\alpha}q^{\alpha}\Big)\Big(n_{\alpha}q^{\alpha}\Big)^{2}w_{2} - 10\Big(q_{\alpha}q^{\alpha}\Big)^{2}w_{2} - 18 \Big(q_{\alpha}q^{\alpha}\Big)\Big(n_{\alpha}q^{\alpha}\Big)^{2}w_{2}^{2} - \Big(q_{\alpha}q^{\alpha}\Big)^{2}w_{2}^{2} \\&&\nonumber
- 8\Big(q_{\alpha}q^{\alpha}\Big)^{2}w_{1}w_{2} + 54\Big(q_{\alpha}q^{\alpha}\Big)\Big(n_{\alpha}q^{\alpha}\Big)^{2}w_{1}w_{2} - 54\Big(n_{\alpha}q^{\alpha}\Big)^{4}w_{1}w_{2} + 27\Big(n_{\alpha}q^{\alpha}\Big)^{4}w_{2}^{2} + 27\Big(n_{\alpha}q^{\alpha}\Big)^{4}w_{1}^{2} - 36\Big(q_{\alpha}q^{\alpha}\Big)\\&&\label{master2}
\Big(n_{\alpha}q^{\alpha}\Big)^{2}w_{1}^{2} + 8\Big(q_{\alpha}q^{\alpha}\Big)^{2}w_{1}^{2}\Big] + 4\Big(q_{\alpha}q^{\alpha}\Big)^{3} = 0.
\end{eqnarray}
\end{widetext}

Eq. (\ref{master2}) can once again be treated as a cubic equation of $\rho_{e}^{2}$. Comparing with the standard formalism given as in Eqs. (\ref{con1}), (\ref{con2}) and (\ref{con3}), we deduce that any or all of the allowed solutions of density can diverge, if and only if the coefficient of highest power of $\rho_{e}$ in Eq. (\ref{master2}) is zero (the term equivalent of $a$ in $ax^{3} + bx^{2} + c x + d = 0$), or
\begin{eqnarray}\nonumber
&& \Big[w_{2}^2 - w_{1}^2 + 2w_{1}w_{2} - 2w_{2}^3 + 2w_{1}^{2}w_{2} + 2w_{2}^{2}w_{1} - 2w_{1}^{3}w_{2} \\&&\nonumber
- w_{1}^{4} - 2w_{1}w_{2}^{3} - w_{2}^{4} + 6w_{1}^{2}w_{2}^{2} + 2w_{1}^{3}w_{2}^{2} + 2w_{2}^{3}w_{1}^{2} - \\&&\nonumber
2w_{1}w_{2}^{4} - 2w_{2}w_{1}^{4} - w_{2}^{4}w_{1}^{2} + 2w_{1}^{3}w_{2}^{3} - w_{1}^{4}w_{2}^{2} - 2w_{1}^{3} \Big] = 0.
\end{eqnarray}    

This gives a general constraint between radial and tangential pressure of the collapsing fluid if the density has to diverge at a singularity. If we put $w_2 = 0$ (zero tangential pressure) the above relation simplifies into $w_{1}^2 (w_{1}^2 + 2w_{1} + 1) = 0$, implying either of the two cases : $w_{1} = 0$, i.e., the interior fluid behaving as an isotropic dust, or $w_{1} = -1$, i.e., the radial pressure component is dark-energy like. A similar deduction can be found for $w_{2}$ by substituting $w_{1} = 0$ (zero radial pressure).  \\

{\bf Case IV : Non-linear equation of state} \\

One can argue that towards the end of a gravitational collapse, with the collapsing shells piling up one after another towards the centre of the sphere, the equation of state or the correlation between pressure components and the density may not be strictly linear. We include a brief note on this; just to point out that the restriction derived from the syzygy remains similar even with a non-linear equation of state. Assuming that
\begin{equation}\label{nonlineos}
p_{r} = \beta_{1}\rho_{e} + \gamma_{1}\rho_{e}^{2} ~,~ p_{t} = \beta_{2}\rho_{e} + \gamma_{2}\rho_{e}^{2},
\end{equation}
we rewrite Eq. (\ref{mastereq}) and identify the highest powers of $\rho_{e}$. In a density-dominated end-state, the three highest powers of density are $\rho_{e}^{12}$, $\rho_{e}^{11}$ and $\rho_{e}^{10}$, making this a cubic equation of $\rho_{e}^{4}$. We have already discussed about the roots of cubic equation earlier in the article and following a similar pathology, we find that the coefficient of $\rho_{e}^{12}$ is $\gamma_{1}^{2}\gamma_{2}^{4} + \gamma_{1}^{4}\gamma_{2}^{2} - 2\gamma_{1}^{3}\gamma_{2}^{3}$. Therefore, one can write the sum of the roots of this cubic equations as
\begin{widetext}
\begin{equation}
\left(\rho_{e1}^{4} + \rho_{e2}^{4} + \rho_{e3}^{4}\right) \simeq \frac{2\gamma_{1}\gamma_{2}\left[\gamma_{1}\gamma_{2}\left(\gamma_{1}+\gamma_{2}\right) - \gamma_{2}^{3}\left(1+\beta_{1}\right) - \gamma_{1}^{3}\left(1+\beta_{2}\right) + \gamma_{1}\gamma_{2}^{2}\left(3\beta_{1} - 2\beta_{2}\right) - \gamma_{1}^{2}\gamma_{2}\left(2\beta_{1} - 3\beta_{2}\right) \right]}{\gamma_{1}^{2}\gamma_{2}^{2}\left(\gamma_{1} - \gamma_{2}\right)^2}.
\end{equation}
\end{widetext}
Thinking in reference with Eqs. (\ref{con1}), (\ref{con2}) and (\ref{con3}) it is easy to derive the fact that $\gamma_{1}^{2}\gamma_{2}^{2}\left(\gamma_{1} - \gamma_{2}\right)^2$ will always be on the denominator while defining the other algebraic equations governing the roots. Any one or all of the roots can diverge at a singularity iff the denominator is zero. This can clearly happen if $\gamma_{1} = \gamma_{2}$. In reference to Eq. (\ref{nonlineos}) this indicates an isotropy in the highest power of $\rho$. Intuitively, this argument can be extended to cases where higher powers of density are included in the equation of state. Therefore, a formation of space-time singularity requires a divergence of density and that can only happen if there is a tendancy of developing pressure isotropy.  \\ 

Curvature singularities are a fundamental aspect of GR. This article highlights a salient feature of such singularities through a simple yet significant observation. The conventional portrayal of a formation of singularity involves a solution to the Einstein field equations. For example, to describe a gravitational collapse : there must be a time-evolving solution implying a collapsing metric tensor and a flow of the matter-energy momentum $T_{\mu\nu}$ components towards an eventual divergence. We argue here that the formation of curvature singularities is not always guaranteed even if $T_{\mu\nu}$ satisfies the field equations. We examine the relationship between $T_{\mu\nu}$ components and the polynomial degeneracies among higher-order Ricci and Weyl invariants. By treating these degeneracies as polynomial equations, we explore the nature of possible roots and the constraints they impose on the density, radial pressure, and tangential pressure of the spherical distribution. Our findings indicate that for a singularity ($\rho \to \infty$) to form, the fluid must evolve into one of three states: (i) a pressure-less dust, (ii) an isotropic fluid or (iii) a distribution with negative pressure. As long as the field equations are satisfied by $T_{\mu\nu}$ components, a generic evolution, for example a gravitational collapse remains entirely possible for any imperfect spherical distribution. However, to evolve into a singularity there is an additional requirement of transition into pressure isotropy and/or dust-like behavior. \\

Notably, we also find that a density divergence is allowed in systems with negative pressure, which suggests a possible connection to dark energy. Although collapsing models involving dark energy are rare \cite{scalarcollapse1, scalarcollapse2, scalarcollapse3, nbsc}, this result opens avenues for exploring the potential clustering of dark energy through gravitational collapse. As far as singularities in cosmological solutions are concerned, we note that that results discussed in this article can potentially be more interesting. It is indeed possible for a spatially homogeneous and isotropic universe to start from a big bang singularity purely due to the fact that for such a geometry the syzygy can provide singular roots. Quite similarly a dust universe or a dark energy driven universe too, can have embedded singularities such as big bang or future singularities. On the contrary, singularities can not be consistently associated with anisotropic cosmological models.

\section*{Acknowledgement}

The author acknowledges the IUCAA for providing facility and support under the visiting associateship program. Acknowledgment is also given to Vellore Institute of Technology for the financial support through its Seed Grant (No. SG20230027), the year 2023.

\end{document}